\documentclass[showpacs,prl,floatfix,twocolumn,amsmath,a4]{revtex4}

\usepackage{graphicx}

\begin{document} 

\title{Theoretical prediction of multiferroicity in double perovskite Y$_2$NiMnO$_6$}

\author{Sanjeev Kumar$^{1,2}$}
\author{Gianluca Giovannetti$^{3}$}
\author{Jeroen van den Brink$^{1,4}$ }
\author{Silvia Picozzi$^3$}
\affiliation{$^1$Institute Lorentz for Theoretical Physics, Leiden University, 2300 RA Leiden, The Netherlands}
\affiliation{$^2$ Faculty of Science and Technology,
University of Twente, P.O. Box 217, 7500 AE Enschede, The Netherlands}
\affiliation{$^3$Consiglio Nazionale delle Ricerche - Istituto Nazionale per la Fisica della Materia (CNR-INFM), CASTI Regional Laboratory, 67100 L'Aquila, Italy}
\affiliation{$^4$ Institute for Molecules and Materials, Radboud Universiteit Nijmegen,
P.O. Box 9010, 6500 GL Nijmegen, The Netherlands}
\begin{abstract}

We put forward double perovskites of the R$_2$NiMnO$_6$ family (with $R$ a rare-earth atom) as a new class of multiferroics on the basis of {\it ab initio} density functional calculations. We show that changing $R$ from La to Y drives the ground-state from ferromagnetic to antiferromagnetic with $\uparrow\uparrow\downarrow\downarrow$ spin patterns. This E$^*$-type ordering breaks inversion symmetry and generates a ferroelectric polarization of few $\mu C/cm^2$. By analyzing a model Hamiltonian we understand the microscopic origin of this transition and show that an external electric field can be used to tune the transition, thus allowing electrical control of the magnetization.
\end{abstract}

\date{\today} 


\maketitle

Materials with simultaneous magnetic and ferroelectric ordering -- multiferroics -- are attracting enormous scientific interest due to their potential for applications in memory and data storage devices \cite{Cheong07,Ramesh07}. Multiferroics with magnetically induced ferroelectric order are particularly interesting due to their strong magneto-electric coupling, which is required for an electric (magnetic) control of magnetic (electric) order parameter: a very desirable property from  a technological point of view \cite{Eerenstein06}. Therefore, the search for new multiferroics with the ferroelectric order driven by the magnetic order is currently a very active and important field of research.

In this letter, we show, by using first-principles density functional theory (DFT) calculations and a model Hamiltonian analysis, that the magnetic order in the double perovskite compounds R$_2$NiMnO$_6$ (RNMO) changes from ferromagnetic for R$=$La and Sm to the E$^*$-type for R$=$Y. The E$^*$-type magnetic structure consists of $\uparrow$-$\uparrow$-$\downarrow$-$\downarrow$ spin chains along the {\em cubic} perovskite-like directions (see Fig.~\ref{fig3}), or equivalently, zig-zag ferromagnetic (FM) spin-chains antiferromagentically coupled in-plane to the neighboring zig-zag chain, with an out-of-plane FM coupling \cite{notaE}. E$^*$-type magnetism breaks inversion symmetry and thus allows a ferroelectric polarization to occur. We will show that indeed Y$_2$NiMnO$_6$ is multiferroic with a electric polarization of few $\mu C/cm^2$. We will show that {\it vice versa} the magnetic transition from ferro to E$^*$-type can be tuned by an external electric field, thus allowing electric control of the ferromagnetic order parameter. 


The double perovskite La$_2$NiMnO$_6$ (LNMO) is a ferromagnetic insulator with a Curie temperature close to room temperature ($T_c$ $\sim$ 280 K). The structure of LNMO changes from rhombohedral (R$\bar{3}$) at high temperature, to monoclinic with P2$_1$/n symmetry group at low temperature \cite{Bull,Asai}. DFT has been successful in understanding the ferromagnetic (FM) insulating behaviour and the dielectric anomalies observed experimentally in LNMO \cite{HenaDas2008,Mater}. We work in the framework of density-functional theory, using the Vienna ab initio simulation package \cite{vasp}, in which  the Kohn-Sham equations are computed using the projector augmented wave method (PAW) and solved describing electronic exchange and correlation in the generalized gradient approximation (PBE) \cite{pbe,parameters}.

\begin{figure}
\centerline{\includegraphics[width=.95\columnwidth,angle=0]{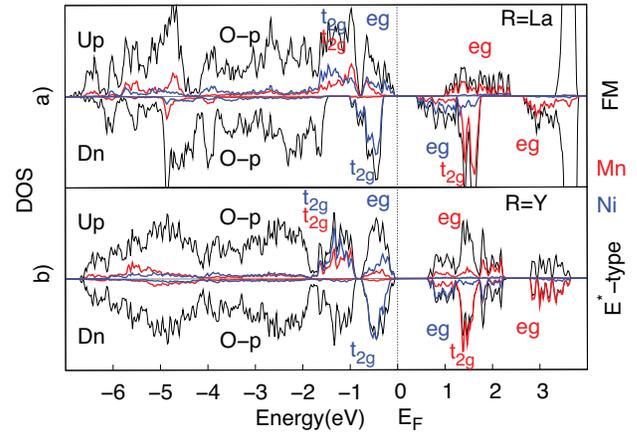}}
\caption{(Color online) (a) Density of states in P2$_1$/n symmetry for FM-type
  magnetic structure in LNMO; (b) Density of states in P2$_1$ symmetry for E$^*$-type
  magnetic structure in YNMO. Zero energy is set to be Fermi level.}
\label{fig1}
\end{figure}

The spin resolved density of states (DOS) of LNMO is shown in Fig. \ref{fig1} a) and they agree well with the one in Ref. \cite{HenaDas2008}. The octahedral surrounding of Mn and Ni atoms split the Mn-d and Ni-d manifolds into the t$_{2g}$ and e$_g$ levels. In the spin up channel the Ni-t$_{2g}$ and Ni-e$_g$ are found within the range of 2 eV below the Fermi level (E$_F$) mixed with O-p and the Mn-t$_{2g}$ states, the latter being localized between the Ni-t$_{2g}$ and the Ni-e$_g$. The Mn-e$_g$ remain empty, separated by $\sim$2.5 eV from the Mn-t$_{2g}$. In the down spin channel the Ni-t$_{2g}$ states are located between the O-p and the Fermi level, while the Ni-e$_g$ lie above the Fermi level along with the Mn-t$_{2g}$ and the Mn-e$_g$. This distribution of states corresponds to the nominal valences Ni$^{2+}$ (d$^8$) and Mn$^{4+}$ (d$^{3}$) : the magnetic moments in our simulations are indeed found to be 1.4 $\mu_b$ and 2.9 $\mu_b$, respectevely, close to the experimental values \cite{Rogado}.

Calculating the DFT total energies of the different magnetic structures on the relaxed crystal structure of LNMO and using a least square mean method to fit their values by an effective Heisenberg Hamiltonian (as in Ref. \cite{Yamauchi}), we extract the magnetic couplings J$_{ij}$ \cite{comment1}.

\begin{table}[h!]
\centering\begin{tabular}{|c|c|c|c|c|c|}
\hline\hline
R & J$_{\parallel (Ni-Ni)}^{NNN}$ & J$_{\parallel (Mn-Mn)}^{NNN}$ & J$_{\perp
  (Ni-Mn)}^{NN}$ &
J$_{\parallel (Ni-Mn)}^{NN}$ & Q \\ \hline
La & -100 & 3 & 44 & 47 & 2.13 \\
Sm & -72  & 3 & 27 & 25 & 2.88  \\
Y  & -57  & 2 & 11 & 12 & 4.75\\ 
\hline\hline
\end{tabular}
\caption{Values of the exchange interactions, J$_{ij}$ (in meV/$\mu_B^2$), for R$_2$NiMnO$_6$  (RNMO) with rare earths R$=$La,Sm,Y: J$_{\parallel (Ni-Ni)}^{NNN}$, J$_{\parallel (Mn-Mn)}^{NNN}$, J$_{\perp  (Ni-Mn)}^{NN}$ and J$_{\parallel (Ni-Mn)}^{NN}$ denote the in-plane Ni-Ni second-nearest neighbor, in-plane Mn-Mn second-nearest neighbor, out-of-plane Ni-Mn nearest-neighbor and in-plane Ni-Mn nearest-neighbor exchange constants, respectively. The  ratio (Q) between J$_{\parallel (Ni-Ni)}^{NNN}$ and J$_{\parallel (Ni-Mn)}^{NN}$ is reported in the last column.}
\label{table1}
\end{table}

Experimentally, the unit cell parameters of double perovskites R$_2$NiMnO$_6$ with R $=$ Sm, Y are unknown. Since the theoretical lattice parameters of LNMO obtained within GGA are very close to the  experimental estimates (see also Ref. \cite{HenaDas2008}), we infer a good accuracy for the prediction of structural properties. Therefore, we determine the lattice parameters and the internal coordinates of the ions for R=Sm, Y by carring out the first-principles volume and shape optimization for the monoclinic phases upon imposing a FM spin configuration. Our results may provide basis to refine the crystallographic structures experimentally \cite{comment12}.  A general trend is found: a smaller size of the radius of the rare earth is such to decrease the Ni-O-Mn angles increasing the octahedral distortions, which in turn reduces the effective
interactions between Ni and Mn ions.  Following this procedure, the total energies of different magnetic structures
as considered above for R=La reveal a transition from a FM (for R=La,Sm) to the E$^*$-type magnetic
structure (for smaller rare-earth ions R=Y) \cite{comment2}.
RNMO with R$=$Sm, Y are found to be insulators with electronic
structures very close to that of LNMO.
We show in Fig. \ref{fig1} b) the spin resolved DOS of
YNMO in the E$^*$-type magnetic structure.
Small differences are found in the values of the band gaps and magnetic
moments at Ni and Mn sites between the considered magnetic structures in all RNMO
studied in this work. 
The magnetism is governed by the superexchange interactions due to
the Hund's rule and energy gain allows virtual hopping of parallel spins and
forbids anti-aligned spins between half-filled Ni-e$_g$ and empty Mn-e$_g$
orbitals as expected in an extended Kugel-Khomskii model \cite{KK,HenaDas2008}.
The trend of the exchange parameters with changing R is reported
in Table \ref{table1}.\cite{comment4}
In-plane exchange interactions between Mn sites are negligible compared to Ni sites.
The ratio between the antiferromagnetic next-nearest-neighbour (NNN), J$_{\parallel (Ni-Ni)}^{NNN}$ and
FM nearest-neighbour (NN) J$_{\parallel (Ni-Mn)}^{NN}$ interactions increases and explains the
stabilization of E$^*$-type magnetic structure:
a frustrated $\uparrow$-$\uparrow$-$\downarrow$-$\downarrow$ Ising--like spin chain in ab
planes \cite{Cheong07,ANNImodel}.
Magnetic couplings changed by intercalation of rare earth
with smaller ionic radius gives the possiblity to achieve a polar state in
RNMO. We recall that the E-type magnetic structure is known to produce large polarization along
the short b axis due to the noncentrosymmetric collinear spin arrangment which breaks the inversion
symmetry in RMnO$_3$ and RNiO$_3$ \cite{picozzi,giovannettiRNiO3,Cheong07,Brink08}.

In order to understand the transition from the FM to the E$^*$-type magnetic state at a more microscopic level, we analyze the following two-band model Hamiltonian for the double perovskites:
\begin{eqnarray}
&& H = -\sum_{ (ij) \sigma}^{\alpha \beta}
t^{ij}_{\alpha \beta} \left ( c^{\dagger}_{i \alpha \sigma} c^{~}_{j \beta \sigma} + H.c. \right ) + \sum_{i} \epsilon_i n_{i} \nonumber \\
&& + U \sum_{i, \alpha} n_{i \alpha \uparrow} n_{i \alpha \downarrow} + (U'+J_H)
\sum_{i} n_{i a \uparrow} n_{i b \downarrow} + n_{i b \uparrow} n_{i a \downarrow} \nonumber \\
&& + (U'-J_H)\sum_{i} n_{i a \uparrow} n_{i b \uparrow} + n_{i a \downarrow} n_{i b \downarrow}
- J_H \sum_{i \in Mn} {\bf S}_i \cdot {\mbox {\boldmath $\sigma$}}_{i} \nonumber \\
&& + \lambda \sum_{i} (Q_{xi} \tau_{xi} + Q_{zi} \tau_{zi} + Q_{bi} n_i) + H_{el}
\end{eqnarray}

Here, $c^{}_{i \alpha \sigma}$ and $c^{\dagger}_{i \alpha \sigma}$ are annihilation and creation
operators for electrons with spin $\sigma$ in the $e_g$ orbital
$\alpha \in \{x^2-y^2 (a), 3z^2-r^2 (b)\}$.
$t^{(ij)}_{\alpha \beta}$ denote the hopping amplitudes between the two
$e_g$ orbitals on NN and NNN sites. The NN Ni-Mn hopping is given by \cite{Slater-Koster}:
$t_{11}^x= t_{11}^y \equiv t$,
$t_{22}^x= t_{22}^y \equiv t/3 $,
$t_{12}^x= t_{21}^x \equiv -t/\sqrt{3} $,
$t_{12}^y= t_{21}^y \equiv t/\sqrt{3} $, where
$x$ and $y$ mark the spatial directions. The NNN hoppings are parameterized by $t'$ such that
$t^{NNN}_{\alpha \beta} = t' ~ t^{NN}_{\alpha \beta}$ \cite{note_hoppings}.
$\epsilon_i$ denotes the on-site energy, and, guided by the electronic
structure shown in Fig \ref{fig1}, 
we set $\epsilon_i(Mn/Ni) = \pm \Delta$. $J_H$ denotes the Hund's rule coupling between the $e_g$ electrons and
$U'$ ($U$) is inter (intra)-band Hubbard repulsion with $U=U'+2J_H$. Since the
Mn sites have a half-filled t$_{2g}$ level, an additional Hund's rule coupling between the Mn-t$_{2g}$ and the Mn-e$_g$ is included.
The ${\mbox {\boldmath $\sigma$}}_i$ denotes the electronic spin operator defined as
${\sigma}^{\mu}_{i}=
\sum_{\sigma \sigma'}^{\alpha} c^{\dagger}_{i\alpha \sigma}
\Gamma^{\mu}_{\sigma \sigma'}
c_{i \alpha \sigma'}$,
where $\Gamma^{\mu}$ are the Pauli matrices. $\lambda$ denotes the strength of the electron-lattice coupling.
$Q_{xi}, Q_{zi}$ are the Jahn-Teller (JT) distortions which are very weak
compared to the breathing mode (BM) distortions $Q_{bi}$. The $\tau_{xi} = \sum_{\sigma} (c^{\dagger}_{i a \sigma}
c^{~}_{i b \sigma} + c^{\dagger}_{i b \sigma} c^{~}_{i a \sigma})$ and
$ \tau_{zi} = \sum_{\sigma} (c^{\dagger}_{i a \sigma}
c^{~}_{i a \sigma} - c^{\dagger}_{i b \sigma} c^{~}_{i b \sigma})$ are the orbital pseudospin operators for the $e_g$ orbitals,
and the elastic energy for the distortions is included as $H_{el} = (1/2)(Q_{xi}^2 + Q_{zi}^2 + Q_{bi}^2)$.
Any deviation from collinear spin arrangements is neglected.

\begin{figure}
\centerline{\includegraphics[width=.95\columnwidth,clip=true]{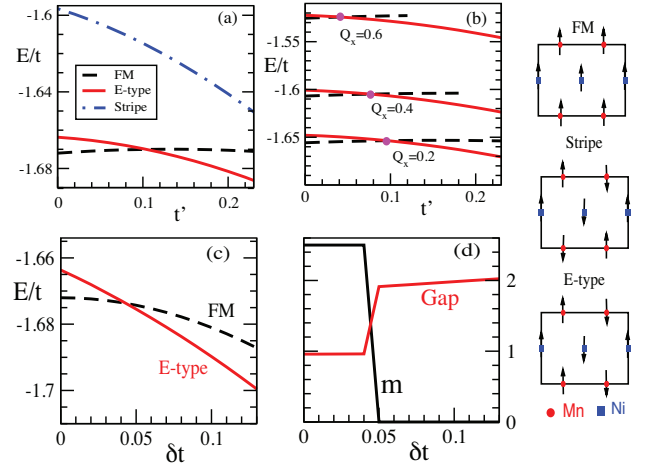}}
\caption{(Color online) Total energy as a function of the ratio $t'$ of the NNN to the NN hopping parameters, (a) in the absence, and (b) in the presence, of JT distortions. The E-type state (equivalent to E$^*$ since the model is 2D) becomes lower in energy with increasing $t'$ in both cases. The other model parameters are $\Delta=2$, $U'=4$, $J_H=0.5 $, $\lambda=1$, $Q_z = 0$, $Q_b = 1$. (c) Total energy as a function of $\delta t$ for $\Delta = 2.0$, $J_H=0.5$, $Q_x=0$ and $U'=4$. (d) Energy gap in the DOS and the net magnetization as a function of $\delta t$.}
\label{fig2}
\end{figure}

\begin{figure}
\centerline{\includegraphics[width=.95\columnwidth,clip=true, angle=0]{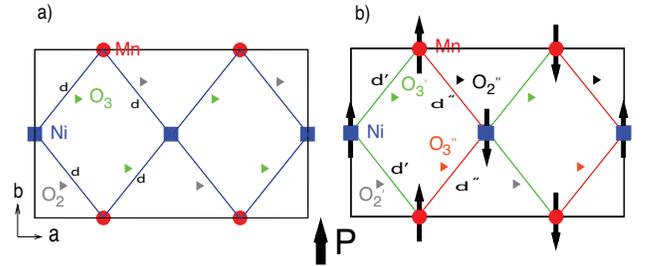}}
\caption{(Color online) In-plane arrangement of the ions and their relative distances in the unit cell  in
P2$_1$/c (a) and  P2$_1$ (b) symmetry. Arrows
indicate the magnetic spin ordering in the E$^*$-type magnetic structure}
\label{fig3}
\end{figure}

We identify the key parameters of the model using inputs from DFT results.
An important feature from DFT calculations is the increase in the relative
strength of the NNN respect to the NN interactions as R changes from La to Y (cfr Table
I). Therefore we focus on this variation, keeping the other model parameters
fixed to reasonable values guided by the DFT results
\cite{HenaDas2008}.
A staggered pattern of the BM distortions with $Q_{bi} = Q_b~e^{{\rm i} (\pi,\pi) \cdot {\bf r}_i}$ is assumed following the DFT results discussed earlier. A weak JT distortion is also considered in order to test the robustness of the results. We analyze the model Hamiltonian by employing a Hartree-Fock decoupling of the interaction terms; replacing $n_{i \alpha \uparrow} n_{i \alpha \downarrow}$ by $\langle n_{i \alpha \uparrow} \rangle n_{i \alpha \downarrow} + n_{i \alpha \uparrow} \langle n_{i \alpha \downarrow} \rangle - \langle n_{i \alpha \uparrow} \rangle  \langle n_{i \alpha \downarrow} \rangle $, etc.
The resulting model is solved self-consistently by iterative numerical diagonalization on a two-dimensional Ni-Mn square lattice starting from different possible magnetic states, and the energies of the converged states are compared to find the groundstate.
In Fig. \ref{fig2} we show the total energy per lattice site as a function of the ratio $t'$ between the NNN and the NN hopping strengths. Panel (a) corresponds to the absence of JT distortions while weak JT distortions are present for the results in panel (b). The figure demonstrates that it is indeed possible to drive a transition from a ferromagnetic to an E$^*$-type state by increasing the NNN hopping parameter.
JT distortions work in favor of the E$^*$-type state as the $t'$ required for the FM to E$^*$-type transition reduces upon increasing $Q_x$ (See Fig. \ref{fig2}b).

Using the Berry-phase(BP) approach we evaluate the first-principles electric polarization P
\cite{berry} for YNMO. The purely electronic contribution to polarization,   P$_{ele}$,
can be calculated by imposing the
E$^*$-type magnetic structure in the centrosymmetric crystal structure with
P2$_1$/n symmetry, leading to  P$_{ele}$=0.75 $\mu C/cm^2$ along the $b$ axis.
The checkerboard pattern due to Ni$^{2+}$ and Mn$^{4+}$ valence in the
monoclinic P2$_1$/n crystal structure leads to the BM
distortions of the NiO$_3$ and MnO$_3$ octahedra,
which contain three inequivalent oxygen atoms (O$_1$ out of plane and O$_2$,
O$_3$ in plane) with
corresponding Ni-O-Mn (Ni-O$_2$-Mn and Ni-O$_3$-Mn) angles differing slightly due to the monoclinic
distortion of the unit cell (see Fig. \ref{fig3} a)).
The E$^*$-type magnetic order further lowers the crystal symmetry to P2$_1$.
The magnetic order couples to the lattice and influences the ionic positions
and the electronic charge distribution, 
leading to a decrease in the electronic contribution with respect to the
purely electronic term calculated for the centrosymmetric crystal structure.
However, the total polarization increases with respect to the centrosymmetric
crystal structure due to the ionic contribution:
P$_{YNMO}$= 2.50 $\mu C/cm^2$ (P$_{ele}$=0.51
$\mu C/cm^2$, P$_{ionic}$=1.99 $\mu C/cm^2$).
The exchange energy in the P2$_1$ symmetry is minimized: (i) by increasing (decreasing)
the distance between $\uparrow$-$\uparrow$ ($\uparrow$-$\downarrow$) along the $\uparrow$-$\uparrow$-$\downarrow$-$\downarrow$
spin chains (see d$^{'}$, d$^{''}$ in Fig. \ref{fig3} b)), which are identical in the original
P2$_1$/n symmetry (see d in Fig. \ref{fig3}); (ii) by further splitting from a
magnetic point of view of the oxygen ions into 4 
inequivalent sites (see Ni-O$_{2^{'}}$-Mn, Ni-O$_{2^{''}}$-Mn and
Ni-O$_{3^{'}}$-Mn, Ni-O$_{3^{''}}$-Mn  in Fig. \ref{fig3} b)).
The inequivalent oxygen atoms O$_{2^{'}}$, O$_{2^{''}}$ and O$_{3^{'}}$,
O$_{3^{''}}$ are charge-polarized with local dipole moments leading to a net
polarization developing along the b axis.
The (2,2) components of Born effective charges (BEC) are consistent with the inequivalency of
oxygen sites: Z$^*_{O_{3^{'}}}$=-2.87 e, Z$^*_{O_{2^{'}}}$=-2.0 e, Z$^*_{O_{3^{''}}}$=-3.65 e,
Z$^*_{O_{2^{''}}}$=-2.76 e (for the labelling of the oxygen sites see Fig. \ref{fig3}).
Within the so called point charge model (PCM) P$_{YNMO}^{PCM}$=1.64 $\mu
C/cm^2$, {\em i.e.}  very close to the one calculated by DFT, confirming that the electronic
rearrangement of the charge due to the symmetry breaking is small.   
Finally, we demonstrate using the model Hamiltonian that an external electric field can induce a magnetic transition in this system. The main effect of an external electric field is to change the on-site energies of the oxygen, which enter in deriving the
hopping parameters used in the model. To leading order
the hoppings between Mn-d and Ni-d via the O-p orbitals are $\propto t_{pd}^2/\Delta_{pd}$. Due to the inequivalence of bridging oxygens( $O_{2^{'}}$ and $O_{2^{''}}$ in Fig. \ref{fig3}), an external electric field makes $\Delta_{pd}$ and hence the d-d hopping parameters staggered. We use a parameter $\delta t$ as a measure of this modulation, such that the hopping in both x and y direction are spatially staggered  $(1 \pm \delta t) ~ t^{NN}_{\alpha \beta}$.
Presence of an electric field (modelled as parameter $\delta t$) leads to a change in the magnetic groundstate of the system from a FM to the E$^*$-type, as seen in Fig. \ref{fig2}c. Fig. \ref{fig2}d shows the discontinuous change in the total magnetization and the gap in the DOS across this transition. The jump observed in the magnitude of the gap is consistent with the DFT values for the two magnetic phases (see fig. \ref{fig1}).

In conclusion, we have performed density functional theory and model calculations to
elucidate the possibility to find novel multiferroics in rare-earths
R$_2$NiMnO$_6$ double-perovskites.
The magnetic state changes from a FM (R=La) to an E$^*$-type (R=Y): the later
being well known to assist in ferroelectricity.
Calculations on a two-band model combined with inputs from DFT calculations describe this magnetic transition.
Y$_2$NiMnO6 is shown to be polar in its magnetic ground state with an intrisic polarization
comparable with other magnetically-driven ferroelectrics ~\cite{Kimura03,Hur04,Giovannetti08,Lottermoser04}.
We suggest experimental investigations to test our theorethical predictions.
We further show that an external electric field can change the magnetic ordering
from FM to E$^*$-type leading to a flip from a non-polar state
with finite magnetization to a polar state with zero net magnetic moment:
a major effect that is of interest for electronic devices.

We thank T. Fukushima and K. Yamauchi for stimulating discussions.
This work is supported by Stichting FOM, NCF and NanoNed.
The research leading to part of these results has received funding from the European
Research Council under the European Community
7$^{th}$ Framework Program (FP7/2007-2013)/ERC
Grant Agreement No. 203523-BISMUTH.
Computational support from CASPUR and SARA Supercomputing Centers
are gratefully acknowledged.


\begin{thebibliography}{99}
\bibitem{Cheong07} S.W. Cheong and M. Mostovoy, Nat. Mater. {\bf 6}, 13 (2007).
\bibitem{Ramesh07} R. Ramesh and N.A. Spaldin, Nat. Mater. {\bf 6}, 21 (2007).
\bibitem{Eerenstein06} W. Eerenstein,  N.D. Mathur, and J.F. Scott, Nature {\bf 442}, 759 (2006).
\bibitem{notaE} Our E$^*$ differ from the conventional E-type by the out-of-plane coupling: FM in E$^*$ and AFM in E.
\bibitem{Bull} C.L. Bull {\it et al.}, J. Phys.  Condens. Matter. {\bf 15}, 4927 (2003).
\bibitem{Asai} K. Asai {\it et al.}, J. Phys. Soc. Jpn. {\bf 47}, 1054 (1978).
\bibitem{HenaDas2008} Hena Das {\it et al.}, Phys. Rev. Lett. {\bf 100}, 186402 (2008).
\bibitem{Mater} S.F. Mater {\it et al.}, J. Magn. Magn. Mater {\bf 308}, 116 (2007).
\bibitem{vasp}  G. Kresse and J. Furthm\"uller,  Phys. Rev. B {\bf 54}, 11169 (1996).
\bibitem{pbe} J.P. Perdew {\it et al.}, Phys. Rev. Lett. {\bf 77}, 3865
  (1996).
\bibitem{parameters} The plane-wave cut-off was chosen as 400 eV. A [3,6,5]  mesh was used for the  Brillouin-zone sampling. In the BP approach, we integrated over strings parallel to $a,b,c$ axes, each of them divided in 8 k-points.
\bibitem{Rogado} N.S. Rogado {\it et al.}, Adv. Mat. {\bf }17, 2225 (2005).
\bibitem{Yamauchi} K. Yamauchi {\it et al.}, Phys. Rev. B {\bf 78}, 014403 (2008);
Y. Zhang, H. J. Xiang, and M.-H. Whangbo, Phys. Rev. B {\bf 79}, 054432 (2009).
\bibitem{comment1} Spin-orbit coupling (neglected here) might change
  slightly the values of J$_{ij}$ (see Ref. \cite{HenaDas2008}).
\bibitem{comment12} The PBE theorethical lattice parameters are for R$=$Sm:
  a$=$5.309\AA, b$=$5.513\AA, c$=$7.574\AA, $\beta=$90.218$^{o}$, for R$=$Y:
a$=$5.239\AA, b$=$5.490\AA, c$=$7.494\AA, $\beta=$90.437$^{o}$.
  Details of the internal parameters of the optimized
  structure will be published elsewhere. 
\bibitem{comment2} In order to avoid any additional parameter which could
  make the discussion of the trends more difficult, we do not employ a GGA+U
  approach. Note that also in rare earth manganites bare GGA
  reproduces the correct magnetic ground states \cite{Yamauchi}.
\bibitem{KK} K.I. Kugel ans D.I. Khomskii, Sov. Phys. Usp. {\bf 25}, 231 (1982).
\bibitem{comment4} The $J_{ij}$ values might be affected by details of exchange-correlation potential (GGA vs LDA, addition or not of Hubbard-like correction, etc). However, trends - of primary interests here - are expected not to be affected.
\bibitem{ANNImodel} Fisher {\it et al.}, Phys. Rev. {\bf 44}, 1502 (1980).
\bibitem{picozzi} S. Picozzi {\it et al.}, Phys. Rev. Lett. {\bf 99}, 227201 (2007).
\bibitem{giovannettiRNiO3} G. Giovannetti {\it et al.}, to be published.
\bibitem{Brink08} J. van den Brink and D. Khomskii, J. Phys. Cond. Matt. {\bf 20},
  434217 (2008).
\bibitem{Slater-Koster} J. C. Slater and G. F. Koster, Phys. Rev. {\bf 94}, 1498 (1954).
\bibitem{note_hoppings} In principle the structure of the NNN hopping matrix is different from that of the NN one, but these details are not important for the effect we are interested in.
\bibitem{berry} R.D. King-Smith and D. Vanderbilt, Phys. Rev. B {\bf 47}, 1651
  (1993); R. Resta, Rev. Mod. Phys {\bf 66}, 899 (1994).
\bibitem{Kimura03} T. Kimura {\it et al.},  Nature {\bf 426}, 55 (2003).
\bibitem{Hur04} N. Hur {\it et al.},  Nature {\bf 429}, 392 (2004).
\bibitem{Giovannetti08} G. Giovannetti and J. van den Brink, Phys. Rev. Lett. {\bf 100}, 227603 (2008).
\bibitem{Lottermoser04} T. Lottermoser {\it et al.}, Nature {\bf 430}, 541-544 (2004).
\end{thebibliography}
\end{document}